\documentclass[11pt]{article}
\usepackage{jheppubmod}
\usepackage{psfrag}
\usepackage{comment}
\usepackage{array}
\usepackage{amssymb}
\usepackage{amsmath}
\usepackage{amsthm}
\usepackage{graphicx}
\usepackage{caption}
\usepackage{float}
\usepackage{dsfont}

\title{Anisotopic inflation with a non-abelian gauge field in Gauss-Bonnet gravity}
\author[a]{Sayantani Lahiri}

\affiliation[a]{ZARM, University of Bremen, \\
Am Falltrum,
28359 Bremen, Germany}
\emailAdd{sayantani.lahiri@gmail.com}

\bigskip

\abstract{ In presence of Gauss-Bonnet corrections, we study anisotropic inflation aided by a massless
$SU(2)$ gauge field where both the gauge field and the 
Gauss-Bonnet term are non-minimally coupled to the inflaton. 
In this scenario, under slow-roll approximations, 
the anisotropic inflation is realized as an attractor solution
with quadratic forms of inflaton potential and Gauss-Bonnet coupling function.
We show that the degree of anisotropy is proportional to the additive combination 
of two slow-roll parameters of the theory.
The anisotropy may become either positive or negative similar to the
non-Gauss-Bonnet framework, a feature of the model for anisotropic inflation supported by a non-abelian gauge field
but the effect of Gauss-Bonnet term further enhances or suppresses the generated anisotropy.}
\begin{document}
\maketitle

\section{Introduction}
The framework of cosmological inflation, associated with a period of accelerated expansion of the early universe
\cite{Guth, Star} not only resolves shortcomings of the standard big bang model but
supports cosmological principle at large scales as well.
At the same time, besides implying spatial flatness of the universe, inflation triggers 
primordial fluctuations which account for the formation of large scale structure of  
the universe \cite{Mukhanov}.
These primordial fluctuations are almost adiabatic, produce a nearly scale-invariant power spectrum,
follow an almost Gaussian distribution and give rise to a statistically isotropic universe. 
Together with homogeneity and spatial flatness, these predictions of inflation have been
confirmed by recent cosmological observations by WMAP and Planck \cite{WMAP, Planck}.\\
It is possible to capture essential features of primordial perturbations 
from temporal, spatial de-Sitter symmetries (\cite{Jiro} and references therein) 
and shift symmetry of the inflaton field.
But it has been observed that not all of present cosmological observations 
are respected by the current inflationary paradigm  
for example, WMAP data \cite{WMAP2} hint a possible scale dependence
of the power spectrum, non-Gaussianity in primordial fluctuations \cite{Juan} and traces of 
statistical anisotropy \cite{Planck3} related to
respective violations of temporal de-Sitter symmetry, shift symmetry and spatial de-Sitter symmetry 
thereby indicating that the universe may not possess exact de-Sitter nature
and thus calling for our attention to focus on fine structures of fluctuations, in 
other words on precision cosmology.\\
Under the influence of high accuracy observational data, several attempts have 
been made \cite{Kar, Dim, Val, Yokoyama, Yamamato,Sheikh1}
for generating statistical anisotropy during inflation.
In a model motivated by supergravity \cite{Watanabe}, 
stable anisotropic inflation is realized with the help of a massless $U(1)$ gauge field where it
is shown that if the back reaction of an abelian vector field on the inflaton dynamics is non-negligible, anisotropy 
persists during slow-roll inflation. 
We shall take this approach in the present work however we will not restrict to the model with the $U(1)$
gauge field.\\
Nevertheless, inflation is believed to occur at energy scale where quantum corrections of gravity are
not ignorable and hence requires quantization of gravity in order
to take into consideration of the effects of quantum gravity in the theory of inflation
near Planck scale. 
 In this direction, the superstring theory provides the most consistent formulation  \cite{Green} of quantum gravity
 involving extra dimensions such that in four dimensions, the low energy limit of the 
 fundamental higher dimensional theory appears as 
 higher order corrections in curvature to the Einstein's gravity, the simplest such correction is the 
 Gauss-Bonnet term \cite{Zwibach,Gross} which gives rise to ghost-free theory in four dimensions. 
 Moreover, Gauss-Bonnet term is the first order correction term of Lovelock theory \cite{Lovelock}, 
the generalized version of Einstein's theory.\\
In four dimensions, 
the Gauss-Bonnet term is topologically invariant and does not alter gravitational equations of motion. 
It only contributes non-trivially to the dynamical equations 
when non-minimally coupled to the scalar field.
In the context of precision cosmology \cite{Pullen}, stable anisotropic inflation in presence of 
Gauss-Bonnet correction \cite{lahiri} has been realized
by taking into account of the back reaction of a massless $U(1)$ gauge field when both the abelian 
field and the Gauss-Bonnet term are
non-minimally coupled to the inflaton field.\\
In the present work, we extend our work to the non-abelian sector with an aim to realize anisotropic inflation
supported by a $SU(2)$ gauge field in presence of Gauss-Bonnet term
to investigate impacts of
self-couplings, gauge components and higher curvature corrections
on anisotropic inflation.
We consider a Yang-Mills field and consider that
both the gauge field and the Gauss-Bonnet term are non-minimally coupled to the inflaton. 
Restricting to the large field inflation model, we have at first numerically analyzed 
principal equations of motions by taking Bianchi-I type metric. 
The numerical analysis shows existence of anisotropic inflation in 
the given set-up.  
Then from the analytical study of the dynamical equations  subjected to slow-roll conditions,
we have determined an expression for estimating the degree of
anisotropy generated in this scenario. 
Since the Gauss-Bonnet term is coupled to the inflaton field, this study
further enables us to compare the generated anisotropy with the non Gauss-Bonnet case \cite{Murata}. 
Finally, we accumulate all the inferences
in the last section.  

\section{Anisotropic inflation supported by a $SU(2)$ gauge field and Gauss-Bonnet correction}
We aim to study anisotropic inflation in presence of Gauss-Bonnet correction with the  
help of a massless non-abelian gauge field non-minimally coupled  
to the inflaton field $\phi$ through the 
gauge coupling function $f(\phi)^2$. 
The Gauss-Bonnet term  is coupled to $\phi$ through the function $\xi(\phi)$.
In the given set-up, let the non-abelian gauge field belong to the $SU(2)$ gauge group, for instance 
we consider the Yang-Mills gauge field described by the following algebra,
\begin{equation}
 [T^a,T^b]=i\epsilon^{abc}T^c
\end{equation}
where $T^{a}=\frac{\sigma^{a}}{2}$ ($a=1,2,3$) are generators of $SU(2)$ algebra
and $\sigma^{a}$ are Pauli's matrices. 
The gauge potential for the Yang-Mills field is defined as $A=A_{\mu}^{a}T^{a} dx^{\mu}$ 
with three gauge components $A^{a} (a=1,2,3)$ corresponding to the 
three generators $T^{a}$ of $SU(2)$ gauge group.
Then with the Gauss-Bonnet term, massless non-abelian gauge field and the inflaton field, 
the gravitational action is given by,
\begin{equation}
 S = \int d^{4}x \,\sqrt{-g} \left[\frac{R}{2 \kappa^2} -\frac{1}{2} \nabla _{\mu} \phi \nabla ^{\mu} \phi
 +\frac{1}{8}\xi(\phi) R^2_{GB}- V(\phi)-\frac{1}{2}f(\phi)^2\,\mbox{tr}(F_{\mu \nu}F^{\mu \nu})\right],   \label{action}
\end{equation}
where $\kappa^2$ is the $4$-dimensional gravitational constant and $V(\phi)$ is the inflaton potential.
The Gauss-Bonnet term is given by, 
\begin{equation}
R^2_{GB} = R_{\mu \nu \rho \beta}R^{\mu \nu \rho \beta}
-4 R_{\mu \nu}R^{\mu \nu}+R^2.
\end{equation}
We mention here that (\ref{action}) is invariant under local $SU(2)$ gauge transformation.  
By varying the action with respect to $g_{\mu \nu}$, the equation of motion is given by
\begin{equation}
\begin{array}{rcl}
 G_{\mu \nu} +\kappa^2 P_{\mu \alpha \nu \beta}\nabla^{\alpha}\nabla^{\beta}\xi & =&
 \kappa^2 \left[ \nabla_{\mu} \phi \nabla_{\nu} \phi -
 \frac{1}{2} g_{\mu \nu} 
 \nabla^{\beta}\phi \nabla_{\beta} \phi 
 -  V(\phi)g_{\mu \nu}\right]\\[2mm]
 &&  + 2 \kappa^2 f(\phi)^2 [\mbox{tr}\left( F_{\mu \alpha} F_{\nu \beta} g^{\alpha \beta}\right)-
 \displaystyle\frac{1}{4} g_{\mu \nu} \mbox{tr}( F_{\alpha \beta} F^{\alpha \beta})],  \label{grav_eqn}
 \end{array}
\end{equation}
where $G_{\mu \nu}$ is the Einstein's tensor, $\nabla_{\mu}$ is the covariant derivative with respect
to the metric $g_{\mu \nu}$. 
The Gauss-Bonnet part in the equation of motion is given by
\begin{equation}
\begin{array}{rcl}
 P_{\mu \alpha \nu \beta}\nabla^{\alpha}\nabla^{\beta}\xi & = & R_{\mu \alpha \nu \beta}\nabla^{\alpha}\nabla^{\beta}\xi
 -\Box \xi R_{\mu \nu} \\[2mm]
 & +& \left(\nabla_{\mu}\nabla^{\alpha}\xi R_{\alpha \nu}+\nabla^{\beta} \nabla_{\nu}\xi R_{\mu \beta}\right)
 -\displaystyle\frac{1}{2}\nabla_{\mu}\nabla_{\nu}\xi \\[2mm]
 & - & \displaystyle\frac{1}{2}\left(2 \nabla^{\alpha}\nabla^{\beta}\xi R_{\alpha \beta}
 -R \Box \xi \right)g_{\mu \nu}.                        \label{GBform}
\end{array}
\end{equation}
The equation of motion of the inflaton field  and that of the gauge field  derived from (\ref{action}) 
are respectively given by,
\begin{eqnarray}
 \Box \phi +\frac{1}{8}\xi'(\phi)R^2_{GB}- V'(\phi)
 -f'(\phi)f(\phi)\mbox{tr}(F_{\mu \nu}F^{\mu \nu}) &= & 0      \label{box_scalar} \\[1mm]
 D_{\alpha}[f(\phi)^2\,F^{\mu \alpha}]&=&0                        \label{gauge}
\end{eqnarray}
where $D_{\alpha}$ is the gauge covariant derivative defined as 
$D_{\alpha}=\nabla_{\alpha}+i g_{Y}[A_{\alpha},..]$ and $'$ denotes derivative with respect to $\phi$.
The field strength for the Yang-Mills field represented by $SU(2)$ algebra is given by 
$F_{\mu \nu}=\partial_{\mu}A_{\nu}-\partial_{\nu}A_{\mu}+i g_{Y}[A_{\mu},A_{\nu}]$ 
where $g_{Y}$ is the Yang-Mills coupling constant. 
In order to establish anisotropic inflation in the present scenario, we
consider following Bianchi-I metric which is given by, \\[1mm]
\begin{equation}
 ds^2 = -dt^2 +e^{2\alpha(t)}\left[ e^{-4\sigma(t)}dx^2 + e^{2\sigma(t)}(dy^2 + dz^2) \right],  \label{metric-aniso}
\end{equation}
where $t$ is the cosmic time,  $e^{\alpha}$ is the isotropic scale factor and $\sigma$ indicates deviation from isotropy.
We now choose a gauge such that the temporal component $A_0$ of the gauge potential satisfies $A_{0}=0$.
The existence of rotational symmetry in the $y-z$ plane  governs the form of the
gauge potential, considered also in the non-Gauss-Bonnet case \cite{Murata}.
Thus we have,
\begin{equation}
 A(x^{\mu})= v_{1}(t)T^1 dx + v_{2}(t)(T^2 dy +T^3 dz),  \label{gauge_pot}
\end{equation}
parametrized by the functions $v_{1}(t)$ and $v_2(t)$. 
Using the form of gauge potential given by (\ref{gauge_pot}), $F_{\mu \nu}$ can be constructed.
Assuming that $\phi=\phi(t)$, the equations of motion
obtained by using (\ref{grav_eqn}) - (\ref{box_scalar}), (\ref{metric-aniso}) and (\ref{gauge_pot}) 
are as follows,
\begin{eqnarray}
 \dot{\alpha}^2-\dot{\sigma}^2 &=& \frac{\kappa^2}{3} \left[V(\phi) +\frac{\dot{\phi}^2}{2}
 - 3 \dot{\xi}(\dot{\alpha}-2 \dot{\sigma})(\dot{\alpha}+ \dot{\sigma})^2\right]  \nonumber \\
&& + \frac{\kappa^2}{6} f(\phi)^2 \left[\dot{v_1}^2 e^{-2\alpha+4\sigma}+2\dot{v_2}^2
 e^{-2\alpha-2\sigma}+2g_{Y}^2 v_{1}^2 v_{2}^2 e^{-4\alpha+2\sigma}+g_{Y}^2 v_{2}^4 e^{-4\alpha-4\sigma} \right] 
 \label{ttcomp} \nonumber
 \\
 [1mm]
 \\
2 \ddot{\alpha}+3(\dot{\alpha}^2+\dot{\sigma}^2)& =& \kappa^2 \left[ V(\phi) -\frac{1}{2}\dot{\phi}^2
-2\dot{\xi}(\dot{\alpha}^3+\dot{\sigma}^3+\dot{\alpha}\ddot{\alpha}-\dot{\sigma}\ddot{\sigma})+
\ddot{\xi}(\dot{\sigma}^2-\dot{\alpha}^2)\right] \nonumber \\
&&-\frac{\kappa^2}{6}f(\phi)^2  \left[\dot{v_1}^2 e^{-2\alpha+4\sigma}+2\dot{v_2}^2
 e^{-2\alpha-2\sigma}+2g_{Y}^2 v_{1}^2 v_{2}^2 e^{-4\alpha+2\sigma}+g_{Y}^2 v_{2}^4 e^{-4\alpha-4\sigma} \right] 
 \label{sca-factor}\nonumber \\  
 [1mm]
\\
 \ddot{\sigma}+ 3\, \dot{\alpha} \dot{\sigma}  &=& - \kappa^2 \dot{\xi}\displaystyle
 \left[\dot{\alpha}(3 \dot{\sigma}^2 + \ddot{\sigma})+\dot{\sigma}\left(\ddot{\alpha}+
 2 \ddot{\sigma}\right)+ 3\dot{\alpha}^2\dot{\sigma}\right]- \kappa^2\ddot{\xi} (\dot{\alpha}\dot{\sigma}+\dot{\sigma}^2)\nonumber \\[2mm]
  && + \frac{\kappa^2}{3} f(\phi)^2 \left[\dot{v_1}^2 e^{-2\alpha+4\sigma}-\dot{v_2}^2
 e^{-2\alpha-2\sigma}-g_{Y}^2 v_{1}^2 v_{2}^2 e^{-4\alpha+2\sigma}+g_{Y}^2v_{2}^4 e^{-4\alpha-4\sigma}\right] 
 \label{aniso-comp}\nonumber
 \\
 [1mm]
 \\
 \ddot{\phi} + 3 \dot{\alpha}\dot{\phi} & =&- V'(\phi)+ 3\xi' (\dot{\alpha}+ \dot{\sigma})\left[\dot{\alpha}^3-
 \dot{\alpha}^2\dot{\sigma}+
 \dot{\alpha}(-2 \dot{\sigma}^2+\ddot{\alpha})-\dot{\sigma}(\ddot{\alpha}+ 2\ddot{\sigma})\right] \nonumber \\[2mm]
&& +f(\phi)f'(\phi)\left[\dot{v_1}^2 e^{-2\alpha+4\sigma}+2\dot{v_2}^2
 e^{-2\alpha-2\sigma}-2 g_{Y}^2 v_{1}^2 v_{2}^2 e^{-4\alpha+2\sigma}-
 g_{Y}^2v_{2}^4 e^{-4\alpha-4\sigma}\right] \label{scalar-comp} \nonumber
 \\
 \end{eqnarray}
 where 'dot' represents derivative with respect to time.
 Here $\dot{\xi}=\xi'(\phi) \dot{\phi}$ and $\ddot{\xi}=\xi''(\phi) \dot{\phi}^2+\xi'(\phi)\ddot{\phi}$.
Now the equations of motion of the gauge field using (\ref{gauge}) - (\ref{gauge_pot}) are given by,
\begin{eqnarray}
 \ddot{v}_1 + 2 \displaystyle \frac{f'}{f}\dot{v}_1 \dot{\phi}+ (\dot{\alpha}+4 \dot{\sigma})\dot{v}_1 +
 2 g_{Y}^2 v_1 v_{2}^2\, e^{-2\alpha-2\sigma}& = &0    \label{v1}  \\[1mm]
 \ddot{v}_2 + 2 \displaystyle \frac{f'}{f}\dot{v}_2 \dot{\phi}+ (\dot{\alpha}-2 \dot{\sigma})\dot{v}_2 +
  g_{Y}^2 v_{1}^2 v_{2}\, e^{-2\alpha + 2\sigma}+ g_{Y}^2 v_{2}^3\, e^{-2\alpha - 2\sigma}& = & 0
  \label{v2}
 \end{eqnarray}
All above equations of motion reduces to the abelian case when $v_2=0, \dot{v}_2=0$. 
The slow-roll inflation is accompanied by approximations namely $\dot{\phi}^2<<V(\phi),\,  
\dot{\xi}\dot{\alpha}<<1$, $\ddot{\xi}<<\dot{\xi}\dot{\alpha}$
and additionally $\alpha>>\sigma$, $\dot{\sigma}<< \dot{\alpha}$ hold, so (\ref{ttcomp}) yields
the Friedmann equation,
\begin{equation}
\dot{\alpha}^2\simeq \frac{\kappa^2}{3}V(\phi)   \label{con_infl1}
\end{equation}
such that a nearly constant inflaton potential gives rise to the accelerated expansion of the universe.
Since additionally $\ddot{\phi}<< 3 \dot{\alpha} \dot{\phi}$ is true in the slow-roll regime,
the scalar field equation becomes,
\begin{equation}
 3 \dot{\alpha} \dot{\phi} + V'(\phi)-3 \xi' \dot{\alpha}^4 \simeq 0  \label{con_infl2}
\end{equation}
The inflation sustains as long as the inflaton potential remains dominant over the
energy density of the Yang-Mills field.
In absence of the gauge field  in the slow-roll regime, anisotropy is absent
and conventional isotropic inflation in presence of Gauss-Bonnet corrections is realized when both 
(\ref{con_infl1}) and (\ref{con_infl2}) are satisfied. 
However, when the non-abelian gauge field is present, its energy density 
increases with the expansion of the Universe while slow-roll conditions still remain intact.  
As a result of the back-reaction of the gauge field, 
anisotropic effects begin to be felt such that the inflaton dynamics is governed by 
(\ref{scalar-comp}) which marks the anisotropic inflationary phase
but as a consequence of its back reaction,
the energy density never exceeds the inflation potential.
Under the approximation $\sigma<<\alpha$, the gauge coupling function is given by \cite{lahiri, Martin},
\begin{equation}
 f(\phi)=e^{-2c \kappa^2 \int \frac{3 V}{-3 V'+\kappa^4 \xi' V^2} d \phi}    \label{f}
\end{equation}
where $c$ is a parameter and we define a quantity $Q=\displaystyle\frac{-3 V'+\kappa^4 \xi' V^2}{3 V}$. \\
The next step towards realizing anisotropic inflation in presence of Gauss-Bonnet corrections
is to specify $V(\phi)$ and $\xi(\phi)$. 
In the present paper, we will consider the large field inflation model with quadratic form of 
Gauss-Bonnet coupling function such that,
\begin{equation}
V(\phi) = \displaystyle \frac{1}{2} m^2 \phi^2,    \qquad \xi(\phi)=\xi_{0}\phi^2   \label{form}
\end{equation}
where $m$ is the mass of the inflaton field and $\xi_{0}$ is the constant parameter of the theory arising
due to Gauss-Bonnet corrections.
Then with the assumed form of $V(\phi)$ and $\xi(\phi)$, (\ref{f}) becomes,
\begin{equation}
f(\phi)= e^{\frac{\sqrt{\frac{3}{2}} \,c \tanh ^{-1}
\left(\frac{\kappa ^2 m \sqrt{\xi _0} \phi ^2}{\sqrt{6}}\right)}{m \sqrt{\xi _0}}}
\end{equation}
\subsection{Numerical study}
It has been observed that in presence of 
Gauss-Bonnet corrections, anisotropic inflationary solutions assisted by a massless
$U(1)$ vector field can be constructed for a large class of potential and coupling functions \cite{lahiri} for which
the vector potential has been taken to be $A_{\mu}=(0,v(t),0,0)$.
However, if the massless vector field obeys $SU(2)$ algebra, the non-linear nature of the equations of motion given by (\ref{grav_eqn})-(\ref{gauge}) pose
difficulties in determining exact scaling solutions. 
Therefore, in order to ascertain the existence of anisotropic inflationary phase,
we shall at first study the phase structure with both $U(1)$ and $SU(2)$ massless gauge fields 
in the context of quadratic inflaton potential given by (\ref{form}) in order to 
locate effects that incur due to both the $SU(2)$ vector field and  the Gauss-Bonnet term.
The corresponding phase space structures are obtained by numerically solving equations of motion 
given by (\ref{grav_eqn})-(\ref{gauge}), where the abelian case is retrieved by substituting $v_2=0=\dot{v}_2$ in 
(\ref{grav_eqn})-(\ref{gauge}) \cite{lahiri}.\\
We started the numerical analysis with very small magnitude of the vector field and
assumed $ \kappa=1, c=2,  m=10^{-5}$, $\xi_{0}=1.45 \times 10^5$. 
The initial conditions are taken to be $\phi_{i} =11.9,\,\dot{\phi_{i}}=10^{-10},\,
\alpha=\sigma=0,\, \dot{\alpha}=4.858 \times 10^{-5},
\dot{\sigma}=10^{-10}$ and the initial condition for $\dot{\alpha}$ is determined using (\ref{ttcomp}).
As $\kappa=1$ is set here, all initial conditions and parameters can be expressed as dimensionless
numbers.
\begin{figure}
\begin{minipage}[b]{0.5\textwidth}
\includegraphics[width=2.6 in]{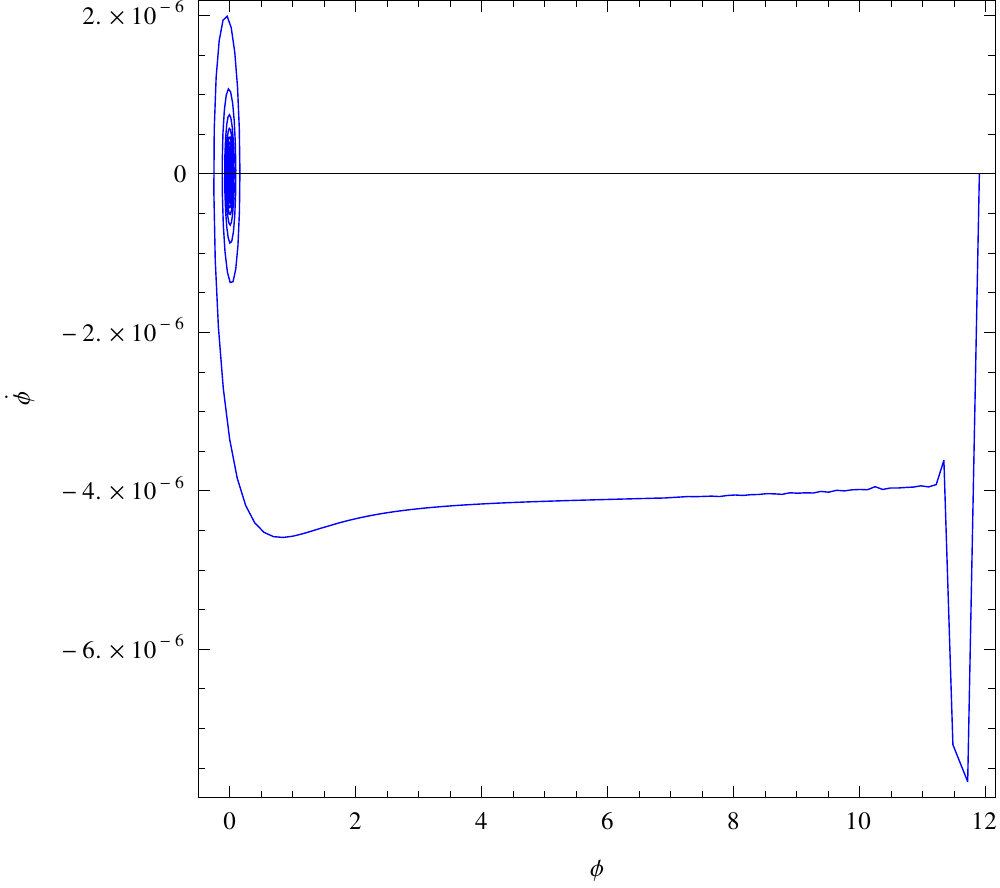}
 \caption*{\\With massless $U(1)$ gauge field.\\ $\xi_{0}=1.45 \times 10^5$, $v=0$ and $\dot{v}=10^{-70}$. }
 \end{minipage} 
   \hfill
   \begin{minipage}[b]{0.5\textwidth}
    \includegraphics[width=2.8 in]{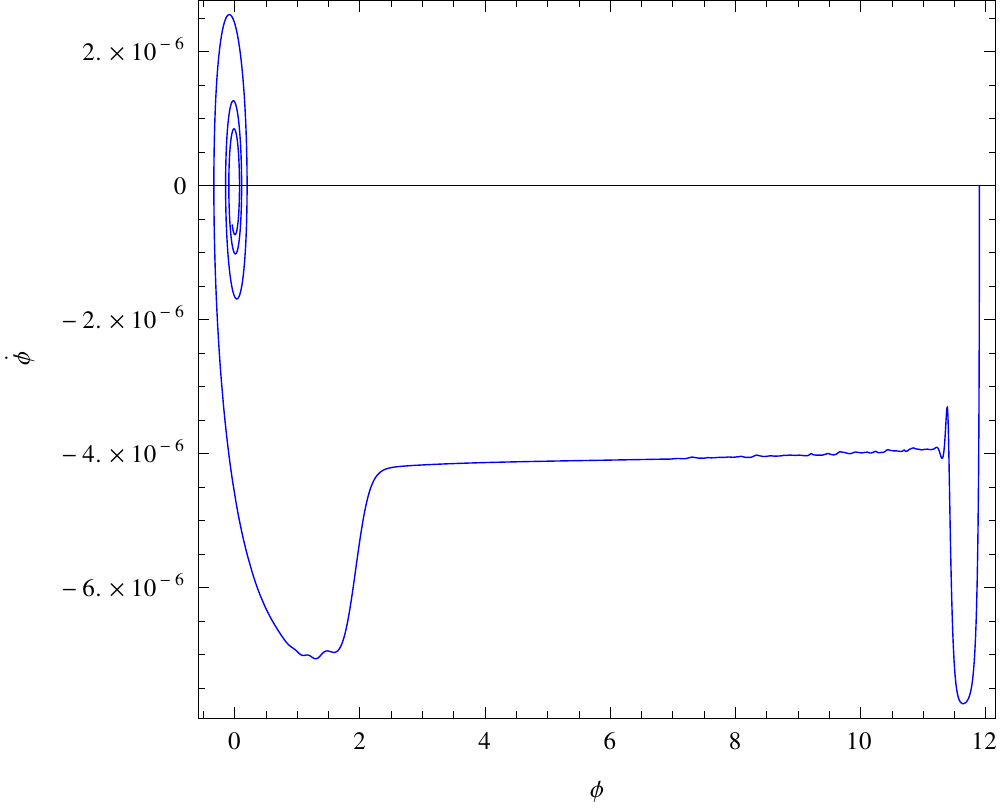}
     \caption*{ With massless $SU(2)$ gauge field.\\ $ \xi_{0}=1.45 \times 10^5, g_{Y}=0.01,v_{1}=v_{2}=0,\\\dot{v}_1 =10^{-70}$ and the ratio
     $\frac{\dot{v_2}}{\dot{v_1}}= 0.6$}
    \end{minipage}
    \caption{The phase flow in $\phi-\dot{\phi}$ space shows existence of isotropic and anisotropic phases of inflation
    triggered by abelian and non-abelian gauge fields. }
\end{figure}
In presence of Gauss-Bonnet corrections, Figure 1 depicts phase flows 
corresponding to abelian and non-abelian gauge fields. 
Both the phase structures are found to be analogous with the inflaton potential, 
Gauss-Bonnet coupling function given by (\ref{form}) and with 
similar choice of parameters and initial conditions.
In this study for the non-abelian case, we assume as an initial configuration that the magnitude of 
$\dot{v}_2$ is proportional to $\dot{v}_1$.
The nature of the phase flow in Figure 1 generated with the help of the $SU(2)$ vector field 
therefore hints to the fact that non-linearity of components of the non-abelian gauge field 
does not significantly contribute to anisotropic inflationary phase
and hence the non-linear terms involving $v_1$ and $v_2$ can be safely neglected.\\
Let us now increase the magnitude of the Gauss-Bonnet parameter further.
Under slow-roll approximations, 
the evolution of slow-roll parameters $\epsilon_{H}$ and $\delta_{H}$ 
(where $\epsilon_v=-\displaystyle\frac{1}{2\kappa^2}\frac{V'}{V}Q$ and 
$\delta_v=\displaystyle\frac{\kappa^2}{3} V' \xi Q$ are their respective counterparts 
in terms of potential and coupling function) in presence of
a non-abelian vector field is obtained numerically using equations of motion  (\ref{grav_eqn})-(\ref{gauge})
as shown in Figure 2. 
In this analysis, initial conditions have been taken as $\phi_{i} =10.5,\,\dot{\phi_{i}}=10^{-10},\,
\alpha=\sigma=0, \,\dot{\alpha}=4.28 \times 10^{-5},\, \dot{\sigma}=10^{-10}$, $v_1=v_2=0$, $\dot{v}_1=10^{-70}$,
$\displaystyle\frac{\dot{v}_2}{\dot{v}_1}=0.6$ and parameters are assumed to be $\kappa=1, c=2, m=10^{-5}, \xi_0= 2.8
\times 10^6$.
From the Figure 2, it is evident that the slow-roll parameter due to Gauss-Bonnet correction 
remains extremely small throughout the inflationary period while
the Hubble's slow-roll parameter approaches unity at the end of inflation.
We note here that the behaviour of slow-roll parameters remain same irrespective 
of the choice of initial conditions.
The Figure 3 depicts the phase flow for Gauss-Bonnet parameter $\xi_0=2.8 \times 10^6$
where both conventional isotropic and anisotropic phases exist. 
The phase structure is obtained under same initial conditions and same 
parameters
considered previously for obtaining plots in Figure 2.\\
\begin{figure}[t]
\begin{minipage}[b]{0.6\textwidth}
 \includegraphics[width=2.0 in]{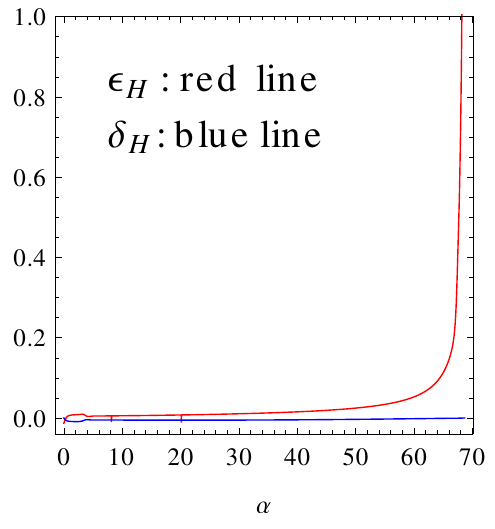}
\end{minipage}
 \hfill
\begin{minipage}[b]{0.6\textwidth}
\includegraphics[width=2.0 in]{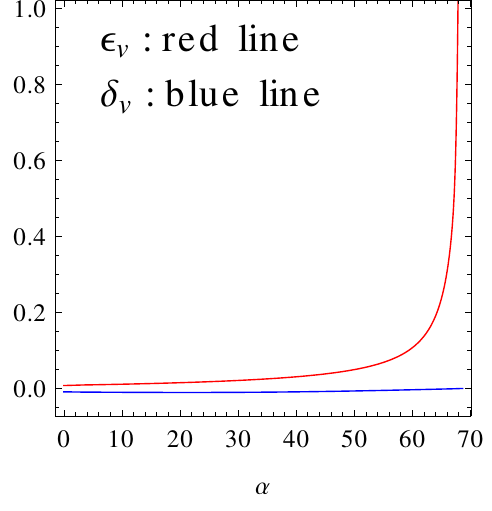}
\end{minipage}
\caption{Plot of slow-roll parameters vs $\alpha$ with $c=2$, $m=10^{-5}$ and $\xi_0=2.8 \times 10^6$.}
\end{figure}
Although the phase structure in Figure 1 shows anisotropic inflation supported by a non-abelian gauge field
is an attractor solution but at the same time the phase flow
depends on the choice of $\dot{v}_2$ \cite{Murata}.
This suggests that in presence of higher curvature corrections, the properties of anisotropic inflation may be examined
by varying the quantity $\displaystyle \frac{\dot{v}_2}{\dot{v}_1}$.
\begin{figure}[h]
\begin{center}
\includegraphics[width=3.4 in]{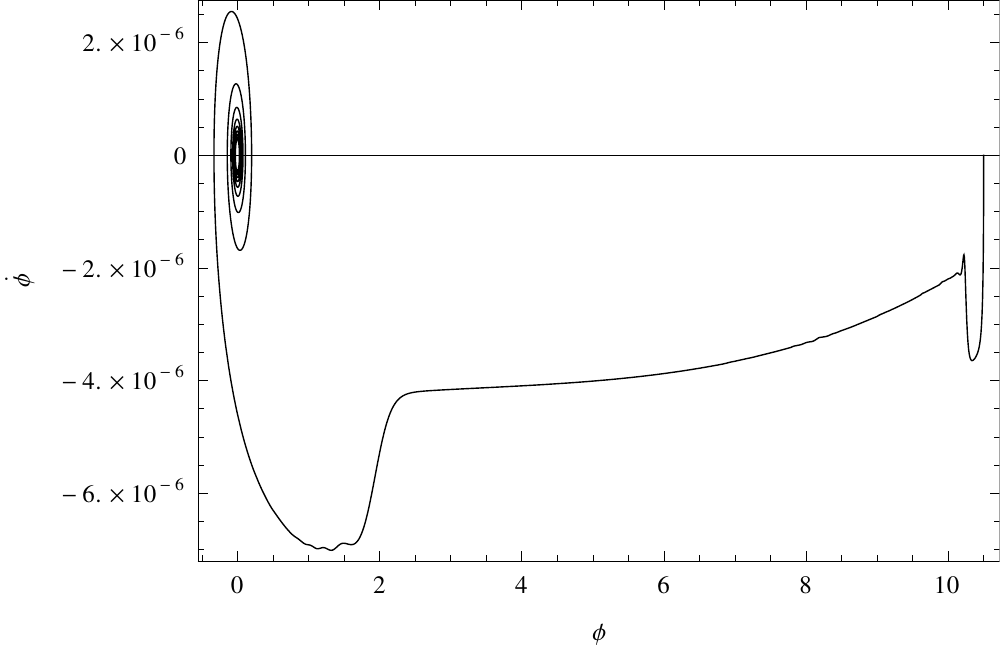}
\end{center}
\caption{The plot of $\phi$ vs $\dot{\phi}$ shows existence of isotropic 
 and anisotropic phases of inflation with initial conditions $\phi_i=10.5$, $\dot{\phi}_i=10^{-10}$
 and parameters $\xi_0=2.8 \times 10^6$, $c=2$, $\displaystyle\frac{\dot{v}_2}{\dot{v}_1}=0.6$.}
 \end{figure}
 In order to study how a gradual increase in $\dot{v}_2$ might affect the evolution of anisotropy 
 $\displaystyle \frac{\Sigma}{H}$ (where $\dot{\sigma}=\Sigma$ with $\dot{\alpha}=H$),
the ratio $\displaystyle\frac{\dot{v}_2}{\dot{v}_1}$ is now slowly varied.
The Figure 4 gives the plot of anisotropy  
vs e-folding number $\alpha$  for different values of $\dot{v}_2$
 \begin{figure}[h!]
 \begin{center}
 \includegraphics{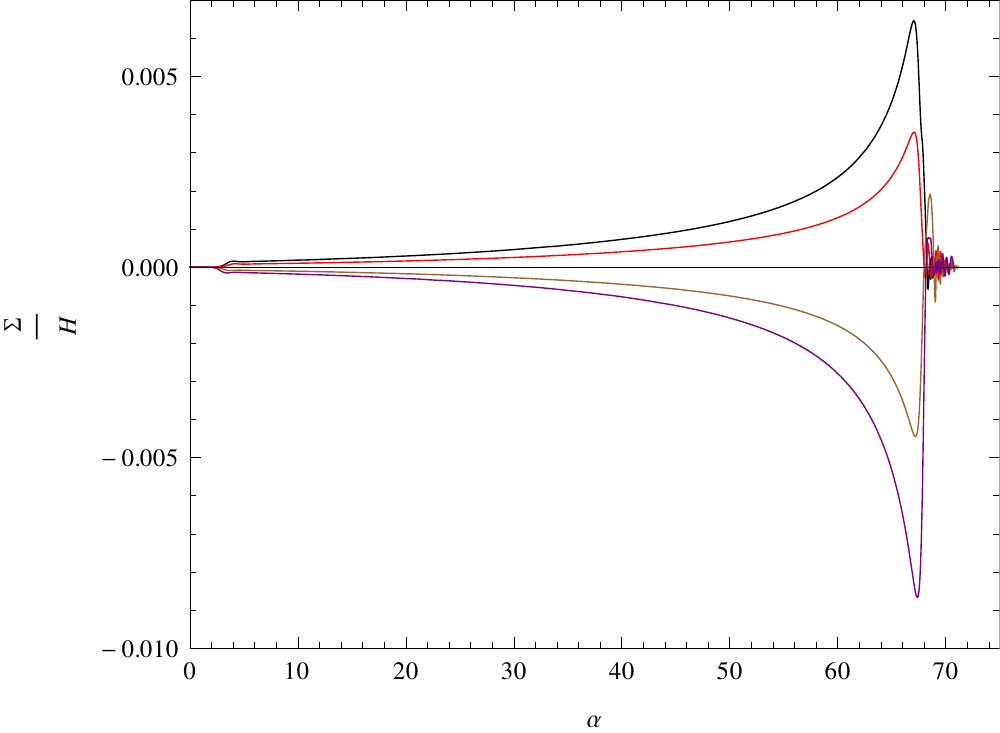}
 \end{center}
 \caption{Plot of $\displaystyle\frac{\Sigma}{H}$ vs $\alpha$ with different values of the ratio
 $\displaystyle\frac{\dot{v}_2}{\dot{v}_1}$.
 In this figure, black, red, brown and magenta plots are produced when 
 $\displaystyle\frac{\dot{v}_2}{\dot{v}_1}$ takes values $0.6, 0.75, 1.5$ and $2.5$ respectively.}
\end{figure}
where it is observed that anisotropy 
generated in presence of a $SU(2)$ vector field and Gauss-Bonnet term is positive 
when the initial configuration of the components of the gauge field obey 
$0 < \displaystyle \frac{\dot{v}_2}{\dot{v}_1}<1$ 
and decreases to zero as the ratio approaches unity and finally 
becomes negative when $\displaystyle \frac{\dot{v}_2}{\dot{v}_1}>1$, 
a feature similar to the corresponding non-Gauss-Bonnet set-up \cite{Murata}.
As shown in Figure 5, the degree of anisotropy 
 $\displaystyle\frac{\Sigma}{H}$ gets enhanced when 
$0 < \displaystyle \frac{\dot{v}_2}{\dot{v}_1}<1$ and becomes more suppressed for
$ \displaystyle \frac{\dot{v}_2}{\dot{v}_1}>1$ compared to non-Gauss-Bonnet scenario. 
We mention here that for obtaining the plot for the evolution of anisotropy corresponding to 
the non-Gauss-Bonnet case in Figure 5, 
 $\phi_i=12, \dot{\phi}_i=0, {v}_1=0={v}_2$ and $\dot{v}_1=10^{-70}$ have been considered.
But this feature is in contrast to the abelian model when
the generated anisotropy is always positive and in particular gets 
enhanced if higher curvature corrections are taken into consideration \cite{lahiri}.
Thus a $SU(2)$ gauge field induced anisotropic inflation 
depends on initial condition of the gauge field in contrary to
the abelian case.
However, as the initial condition dependence appears only in the measurement of anisotropy,
such dependence can be absorbed by rescaling the parameters of the model.\\
Thus the numerical study presented here suggests that even when higher curvature correction like Gauss-Bonnet
term is considered, the non-linearity of the non-abelian vector field does not
affect the nature of anisotropic inflation. 
This inference will be helpful for the analytical study of anisotropic inflation with Gauss-Bonnet correction
presented in the next section.  
\begin{figure}[h]
\begin{minipage}[b]{0.3\textwidth}
 \includegraphics[width=2.5 in]{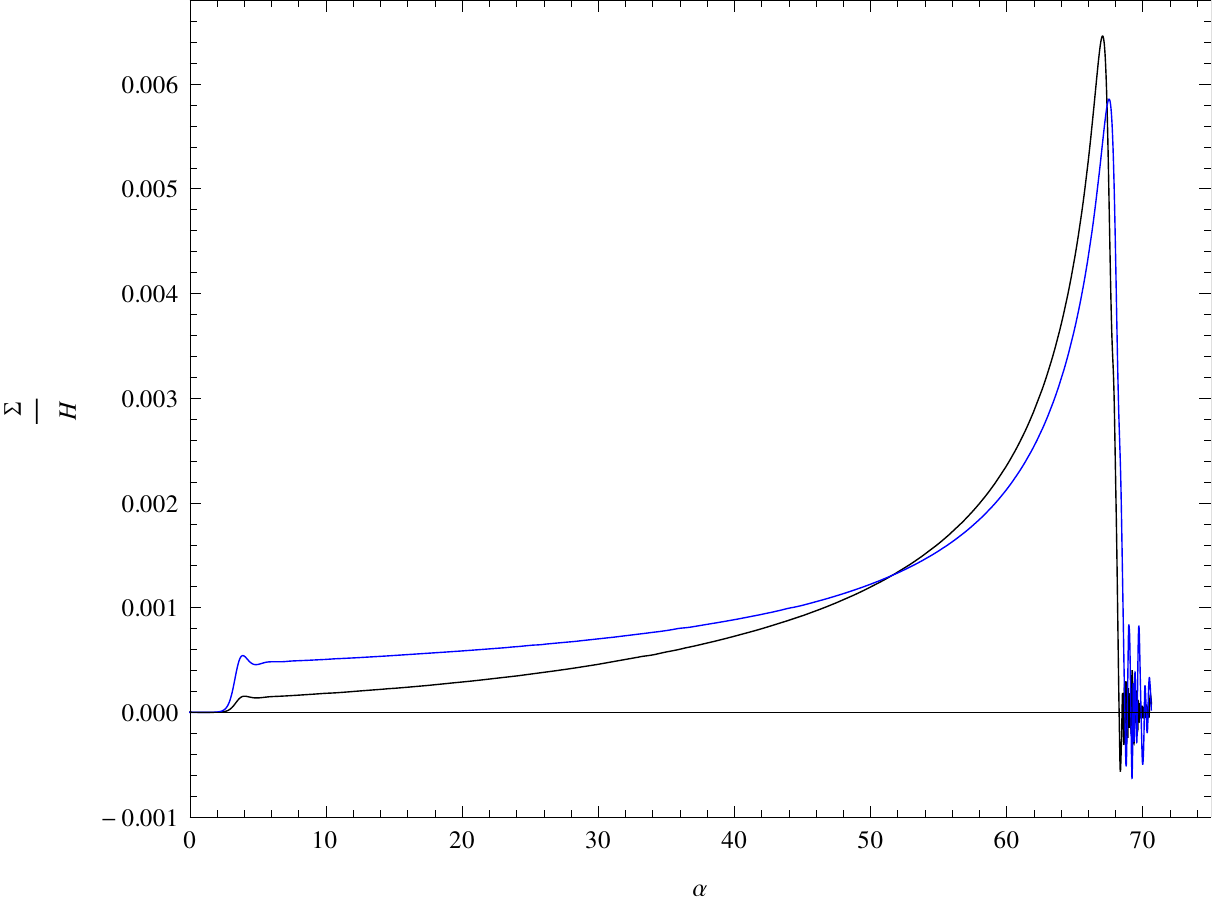}
 \caption*{ $\xi_0=2.8 \times 10^6, \frac{\dot{v}_2}{\dot{v}_1} = 0.6$}
\end{minipage}
 \hfill
\begin{minipage}[b]{0.5\textwidth}
\includegraphics[width=2.5 in]{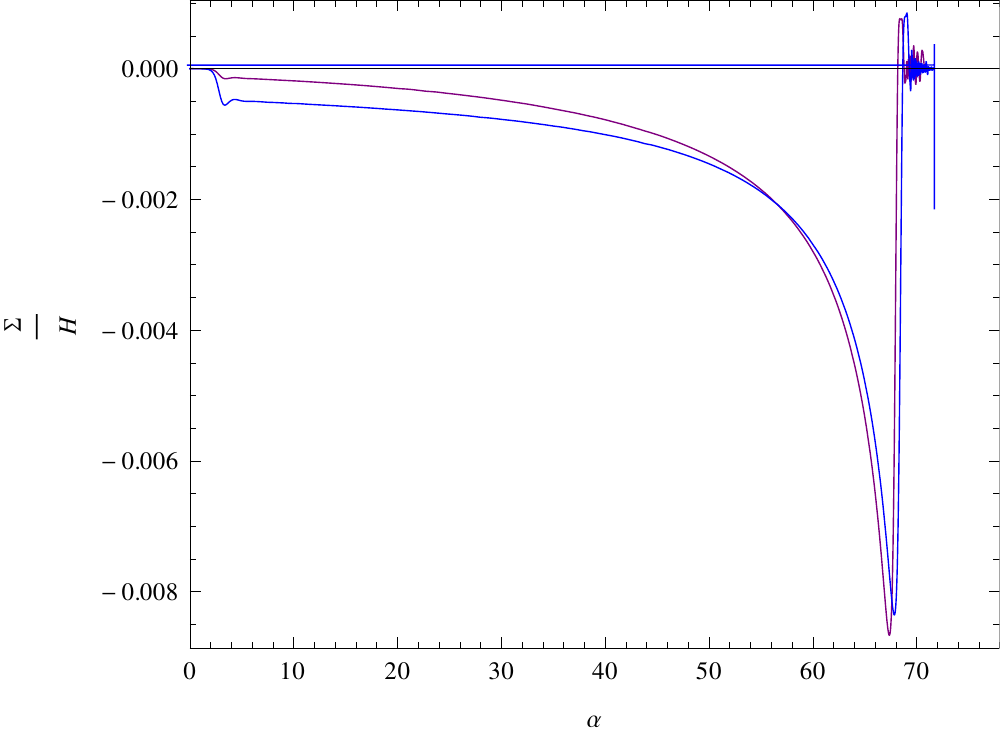}
\caption*{$\xi_0=2.8 \times 10^6, \frac{\dot{v}_2}{\dot{v}_1} = 2.5$.}
\end{minipage}
\caption{Plot of $\displaystyle\frac{\Sigma}{H}$ vs e-folding  number $\alpha$ for comparing the degree of 
anisotropy in Gauss-Bonnet and in non-Gauss-Bonnet set-ups.
Here blue lined plots signify non-Gauss-Bonnet case where $\kappa=1, c=2$, $m=10^{-5}$ are assumed.}
\end{figure}

\subsection{Analytic study} 
During the slow-roll inflationary phase with the non-zero Gauss-Bonnet correction, 
the inflaton field takes the initial value as $\phi_i \sim 10$ for the Gauss-Bonnet parameter $\xi_0=2.8 \times 10^6$, 
so that using (\ref{f}), the gauge coupling function becomes $f(\phi)\sim 10^{62}$. 
But from the action (\ref{action}),
$\displaystyle\frac{g_{Y}}{f(\phi)}$ turns out to be the effective gauge coupling,
that goes as $\displaystyle \frac{g_{Y}}{f(\phi)}\sim 10^{-62}$ which 
is a minuscule quantity indicating that its effect can be ignored in generating anisotropic signatures.
This is corroborated by the numerical analysis which shows that contributions of non-linear terms fade 
away and the anisotropic inflation is an attractor solution. 
These observations indicate that the 
Yang-Mills gauge coupling can be neglected while analytically 
studying equations of motion.
Then equation of motion of the gauge field given by (\ref{v1}) and (\ref{v2}) can be integrated to obtain,
\begin{eqnarray}
 \dot{v}_1=\displaystyle \frac{e^{-\alpha-4\sigma}}{f(\phi)^2}c_1   \label{solv1}\\[1mm]
  \dot{v}_2=\displaystyle \frac{e^{-\alpha+2\sigma}}{f(\phi)^2}c_2  \label{solv2}
\end{eqnarray}
where $c_1$ and $c_2$ are constants of integration. \\
Now, the energy density of the vector field under the condition $\sigma<<\alpha$ becomes,
\begin{equation}
 \rho_{v}=\displaystyle \frac{\kappa^2}{2}f(\phi)^2 e^{-4\alpha}(c_1^2+c_2^2)
\end{equation}
which suggests  at least $f(\phi)=e^{2\alpha}$ is required to commence anisotropic inflation.
More generally, we can parametrize $f(\phi)$ such that
\begin{equation}
 f(\phi)=e^{-2 c \alpha}
\end{equation}
so that $\rho_{v} \propto e^{4(c-1)\alpha}$ which implies $c>1$ in order that $\rho_v$ evolves due to expansion and 
anisotropic effects do not get diluted during the slow-roll regime. 
Then with the condition $c>1$ and using (\ref{f}),  $f(\phi)$, $V(\phi)$ and
$\xi(\phi)$ obey the following relation,
\begin{equation}
\displaystyle \frac{f'(3V'-\kappa^4 \xi' V^2)}{\kappa^2 V f} > 6   \label{rel}
\end{equation}
The above relation suggests that in Gauss-Bonnet set-up, the anisotropic effects during slow-roll inflation can be captured
for a given class of $V(\phi)$, $\xi(\phi)$ provided (\ref{rel}) is satisfied.
It is to be noted as long as $c>1$, the attractor solution exists and therefore the anisotropic phase exists
independent of the choice of a particular value of $c$.\\ 
We will now employ slow-roll approximations in
order to estimate $\displaystyle \frac{\Sigma}{H}$ in presence of higher curvature corrections.
In the entire slow-roll inflationary phase, the energy density $\rho_{v}$
and Gauss-Bonnet contributions remain sub-dominant compared to the inflaton potential 
such that for an almost flat potential profile the universe undergoes an accelerated expansion 
which suggests $\sigma<<\alpha$. 
But as $\rho_v$ grows with expansion, the equation of motion of the inflaton field subjected to slow-roll conditions given by 
$\ddot{\phi}<<3 H \dot{\phi}, \dot{H}<<H^2$, shows up anisotropic effects.
With the approximation $\sigma<<\alpha$ and neglecting higher powers of $\displaystyle\frac{\Sigma}{H}$, 
the scalar field equation after substituting (\ref{solv1}) and (\ref{solv2}) becomes,
\begin{equation}
3 H \dot{\phi}=-V'(\phi)+3 \xi' H^4+\displaystyle \frac{f'(\phi)}{f(\phi)^3}(c_1^2+2c_2^2)e^{-4\alpha} 
\end{equation}
Dividing  the above relation by $3H^2$ and using (\ref{con_infl1}) and (\ref{f}), we obtain,
\begin{equation}
 \displaystyle \frac{d\phi}{d\alpha} = \displaystyle\frac{-3V'+\kappa^4\xi'V^2}{3 \kappa^2 V} +
 \displaystyle \frac{6 c\,(c_1^2+ 2c_2^2)}{\left(3 V'-\kappa^4\xi'V^2 \right)}
 e^{-4\alpha -4\kappa^2 c\int\frac{3 V}{3V'-\kappa^4\xi'V^2} d \phi }   \label{above1}
\end{equation}
Neglecting the variations of $V(\phi)$, $V'(\phi)$ with respect to $\alpha$, integration of (\ref{above1}) gives,
\begin{equation}
 e^{4\alpha +4\kappa^2 c\int\frac{3 V}{3V'-\kappa^4\xi'V^2} d \phi} =
\displaystyle \frac{6 c^2 (c_1^2+ 2c_2^2)}{c-1}\displaystyle \frac{3 \kappa^2 V}{(3V'-\kappa^4\xi'V^2)^2}
(1+A e^{-4(c-1)\alpha})   \label{soln} 
\end{equation}
where $A$ is the constant of integration. Then using (\ref{soln}) we have,
\begin{equation}
 \displaystyle \frac{d\phi}{d\alpha}\,=\,\frac{-3V'+ \kappa^4 \xi' V^2}{3 \kappa^2 V} 
 \left(1-\displaystyle \frac{c-1}{c}[1+ A e^{-4(c-1)\alpha}]^{-1}\right)
\end{equation}
The constant of integration $A$ is fixed by using boundary conditions corresponding to $\alpha \rightarrow \pm \infty$.
\begin{itemize}
 \item $\alpha \rightarrow -\infty$ implies $[1+ A e^{-4(c-1)\alpha}]^{-1} \rightarrow 0$, then
\begin{equation}
 \displaystyle \frac{d\phi}{d\alpha}= \displaystyle \frac{Q}{\kappa^2}
\end{equation}
 which is the conventional slow-roll regime.
\item On the other hand, $\alpha \rightarrow \infty$ implies $[1+ A e^{-4(c-1)\alpha}]^{-1} \rightarrow 1$, so that
\begin{equation}
\displaystyle \frac{d\phi}{d\alpha}= \displaystyle \frac{1}{c}\frac{Q}{\kappa^2}    \label{mod_infl}
\end{equation}
which is the modified slow-roll regime signifying the anisotropic inflation.
\end{itemize}
As Universe expands during anisotropic inflation, 
the anisotropy almost attains a constant value such that $\ddot{\sigma}=\dot{\Sigma}\approx 0 $ and 
$\ddot{\sigma}<< \dot{\alpha} \dot{\sigma}$.
Then using (\ref{con_infl1}), (\ref{f}), (\ref{soln}) and taking into account of slow-roll approximations given by
$\dot{\xi} \dot{\alpha}<<1,\ddot{\xi}<< \dot{\xi} \dot{\alpha}$ and $\sigma << \alpha$, the anisotropy equation
given by (\ref{aniso-comp}) becomes,
\begin{equation}
 \displaystyle \frac{\Sigma}{H}(1+\delta_{H}) =\displaystyle \frac{c-1}{6c^2} 
 \displaystyle \left(\frac{c_1^2-c_2^2}{c_1^2+2 c_2^2}\right) \frac{Q^2}{\kappa^2 }  \label{new-aniso}
\end{equation}
where slow-roll parameter associated with Gauss-Bonnet correction is defined as 
$\delta_{H}=\kappa^2 \dot{\xi} H$ and since $\displaystyle \frac{\Sigma}{H}<<1$, all higher powers of
$\displaystyle \frac{\Sigma}{H}$ are neglected in obtaining (\ref{new-aniso}).
Under slow-roll conditions and with $\sigma << \alpha $, $\dot{\Sigma}=0$, substitution of (\ref{ttcomp}), (\ref{f}),
(\ref{soln}) in (\ref{sca-factor}) yields,
\begin{equation}
 \ddot{\alpha}=-\displaystyle\frac{1}{2}\kappa^2\dot{\phi}^2 + \displaystyle \frac{1}{2}\kappa^2 \dot{\xi} \dot{\alpha}^3
 -\displaystyle  \frac{c-1}{6c^2} \frac{(3V-\kappa^4 \xi'V^2)^2}{9 V}        \label{above2}
\end{equation}
It is known that the Hubble's slow-roll parameter is given by 
$\epsilon_{H}=-\displaystyle \frac{\ddot{\alpha}}{\dot{\alpha}^2}$. 
Now, dividing (\ref{above2}) by $(-\dot{\alpha}^2)$ and combining it with (\ref{con_infl1}) and (\ref{mod_infl}) gives,
\begin{equation}
 \epsilon_{H}+\frac{1}{2}\delta_{H}=\frac{1}{c}(\epsilon_v+ \frac{1}{2}\delta_v)   \label{above3}
\end{equation}
where slow-roll parameters expressed in terms of inflaton potential and Gauss-Bonnet coupling
function are given by
$\epsilon_v=-\displaystyle\frac{1}{2\kappa^2}\left(\frac{V'}{V}\right)Q$ and
$\delta_v=\displaystyle\frac{1}{3} \kappa^2 \xi' V Q$.
But in  particular $\delta_{H}<<1$  
so substituting (\ref{above3}) back in (\ref{new-aniso}) 
gives the measure the anisotropy in presence of Gauss-Bonnet correction and a non-abelian gauge field as,
\begin{equation}
 \displaystyle \frac{\Sigma}{H}=\displaystyle \left(\frac{c_1^2-c_2^2}{c_1^2+2 c_2^2}\right) 
 \displaystyle \frac{c-1}{3c}(\epsilon_{H}+\frac{1}{2}\delta_{H})  \label{aniso_nab}
\end{equation}
which is found to be proportional to slow-roll parameters namely $\epsilon_H$
and $\delta_H$ similar to the abelian case \cite{lahiri}, additionally 
the imprint of the $SU(2)$ gauge field appears through the constant $c_2$. 
In particular, the measure of anisotropy during anisotropic inflation which is
triggered by an abelian gauge field in presence of the Gauss-Bonnet term, 
is retrieved by putting $c_2=0$. 
But $\displaystyle \frac{\Sigma}{H}$ exactly 
vanishes when $c_1=c_2$, a situation which is not useful for our study.
Since $c>1$ is required to hold for anisotropic effects to persist, it is seen from (\ref{aniso_nab}) that
$\displaystyle \frac{\Sigma}{H}$
becomes negative if $c_2>c_1$ or equivalently $\displaystyle\frac{\dot{v_2}}{\dot{v_1}}>1$ as 
observed in Figure 4. 
This feature is inherent to the model of anisotropic inflation induced by an non-abelian vector field.
In absence of Gauss-Bonnet corrections i,e when  $\delta_H=0$, (\ref{aniso_nab})
reduces to the result obtained in \cite{Murata}.
\subsubsection*{ Determination of $\phi_i$ during anisotropic inflation}
We now determine the initial value of the inflaton field $\phi_i$ governing the
phase flow which depends on parameters of the theory namely $m, c$ and $\xi_0$. 
At the end of the inflation, we have $\epsilon_H=1$. 
Then (\ref{above3}) yields,
\begin{equation}
 \delta_H=\displaystyle \frac{2}{c} \left(\epsilon_v+ \frac{1}{2}\delta_v\right)-2  \label{above4}
\end{equation}
With the quadratic form of potential given by (\ref{form}), we obtain
\begin{equation}
 \delta_v=\displaystyle \frac{1}{3}m^2 \xi_0 \phi^3 \left(-\frac{2}{\phi}+\frac{1}{3}m^2 \xi_0 \phi^3 \right)
\end{equation}
where $\kappa=1$ is taken.
At the end of slow-roll inflation, the inflaton field settles for a small value such that
$\phi \sim \mathcal{O}(1)$ which suggests
$\delta_v$ will be very small provided $m^2\xi_0<<1$. 
Also $\delta_H$ is very small compared to $\epsilon_H$ and $\epsilon_v$ so that from (\ref{above3}) and
(\ref{above4}) we obtain,
\begin{equation}
 \epsilon_H=\displaystyle \frac{2}{c}\epsilon_v=1 \label{above5}
\end{equation}
so long as $m^2\xi_0<<1$ holds. Using (\ref{form}) we can express (\ref{above5}) as,
\begin{equation}
 \frac{2}{\phi ^2}-\frac{1}{3} m^2 \xi _0 \phi ^2-c=0    \label{eqnphif}
\end{equation}
which is a forth-order polynomial equation in $\phi$.
The solution of (\ref{eqnphif}) gives the value of the inflaton at the end of inflation denoted by $\phi_f$.
Now on solving (\ref{eqnphif}) gives two positive and two negative real roots of $\phi$.
Discarding the negative value of the inflaton field, the positive roots are given by,
\begin{equation}
 \phi_1=\frac{\sqrt{-\displaystyle\frac{\sqrt{9 c^2+24 m^2 \xi _0}+3 c}{m^2 \xi _0}}}{\sqrt{2}},\qquad 
  \phi_2=\frac{\sqrt{\displaystyle\frac{\sqrt{9 c^2+24 m^2 \xi _0}-3 c}{m^2 \xi _0}}}{\sqrt{2}}
\end{equation}
where $c>1$ is required for the commencement of anisotropic phase of inflation.\\
Let us at first assume the case when Gauss-Bonnet parameter is positive i.e. $\xi_0>0$, then,
\begin{equation}
\phi_f = \phi_2= \frac{\sqrt{\displaystyle\frac{\sqrt{9 c^2+24 m^2 \xi _0}-3 c}{m^2 \xi _0}}}{\sqrt{2}}  \label{phif}
\end{equation}
which depends on parameters $c, \,\xi_0$ and $m$.
From (\ref{phif}), we find that $\phi_f$ is real and non-zero
provided $\displaystyle \frac{4 m^2 \xi_0}{3c^2} >0$.
We note that the initial value of the inflaton field $\phi_i$ also depends on $c$. 
Using (\ref{mod_infl}) which is valid during modified slow-roll phase, $\phi_i$ is determined from the
e-folding number which is now given by,
\begin{equation}
 N\simeq \int_{\phi_i} ^{\phi_f}\displaystyle \frac{3Vc}{-3V'+\kappa^4\xi'V^2} d\phi
\end{equation}
Assuming the e-folding number $N \approx 71.5$, the above relation using (\ref{form}) yields,
\begin{equation}
 N\simeq \int_{\phi_i} ^{\phi_f}\displaystyle \frac{3 c \phi }{m^2 \xi _0 \phi ^4-6} \approx 71.5  \label{valphi1}
\end{equation}
Using $\phi_f$ from (\ref{phif}), $\phi_i$ is determined by evaluating (\ref{valphi1}) 
for specific values of $c_0$ $m$ and a positive value of $\xi_0$.
In the present case, we have assumed $\kappa=1, c=2$, $m=10^{-5}$ and $\xi_0=2.8 \times 10^6$ for which $\phi_i=10.5$ is obtained
where with our given choice of parameters $m^2\xi_0\simeq 10^{-4}<<1$.
In fig 6, the variation of slow-roll parameter $\delta_v$ evaluated at any $\phi_i$ vs Gauss-Bonnet 
parameter $\xi_0$ shows that for $m=10^{-5}$, $c=2$ and under the condition $m^2\xi_0<<1$, the effect of Gauss-Bonnet 
contributions on the anisotropic inflation reduces as $\xi_0 > 10^7$ and finally diminishes to zero when
$\xi_0$ is increased further. 
With this observation in mind, the Gauss-Bonnet parameter $\xi_0 \sim 10^6$ is considered in our analysis.    
We emphasize here that in absence of Gauss-Bonnet corrections, a similar calculation 
yields $\phi_i=12$ as considered in \cite{Murata}.\\
\begin{figure}[t]
 \begin{center}
 \includegraphics{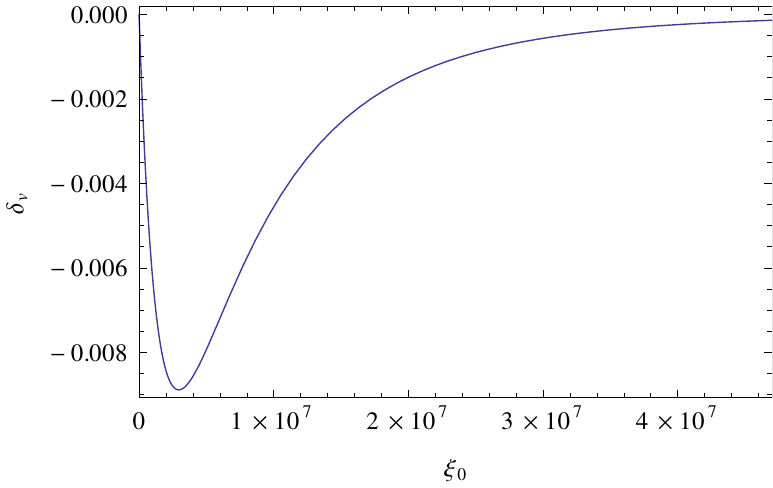}
 \caption{Plot of $\xi_0$ vs slow-roll parameter $\delta_v$ for $\kappa =1, c=2$ and $m=10^{-5}$. }
 \end{center}
\end{figure}
On the other hand, the negative Gauss-Bonnet parameter $\xi_0 <0$ implies $\phi_f=\phi_1$.
Then with $\phi_f=\phi_1$ and $m=10^{-5}$, the initial value of the inflaton field $\phi_i$ evaluated using (\ref{valphi1})
becomes imaginary for large number of values of the parameter $c$.
Hence the negative value of Gauss-Bonnet parameter is discarded in this study.

\section{Concluding remarks }
In the present work we have demonstrated that in presence of Gauss-Bonnet corrections
where the Gauss-Bonnet term is non-minimally coupled to the inflaton field,
a massless non-abelian $SU(2)$ vector field also coupled on-minimally to the inflaton field gives rise to
anisotropic inflation.
In the context of quadratic forms of inflaton potential and Gauss-Bonnet coupling function,
the phase structure obtained under slow-roll approximations, shows existence
of both conventional isotropic and anisotropic phases of inflation under the condition $m^2\xi_0<<1$ 
where $m$ is the mass of the inflaton field and $\xi_0$ is the Gauss-Bonnet parameter.
In the given set-up, the gauge coupling function is determined for the
quadratic form of the inflaton potential and gauss-Bonnet gauge coupling function. 
It is found that anisotropic inflation is an attractor for positive value of 
Gauss-Bonnet coupling parameter $\xi_0$, however 
the attractor solution depends on the adjustment of initial configuration of
one parameter of the $SU(2)$ gauge field.\\
In presence of higher curvature corrections, we have obtained a general relation for 
the measure of anisotropy $\displaystyle \frac{\Sigma}{H}$ which is found to be
proportional to slow-roll parameters namely $\epsilon_H$ and $\delta_H$
but may become either positive 
or negative depending on the initial choice of $\displaystyle \frac{\dot{v_2}}{\dot{v_1}}$. 
This feature, unlike the model of anisotropic inflation with abelian vector field in the
Gauss-Bonnet framework, is facet 
of the existence of multiple components of the non-abelian gauge field.
In particular, we observe that due to the Gauss-Bonnet correction, the slow-roll parameter $\delta_H$ 
either enhances the anisotropy for $\displaystyle \frac{\dot{v}_2}{\dot{v}_1}<1$
or suppresses it more in case $\displaystyle \frac{\dot{v}_2}{\dot{v}_1}>1$ 
compared to the non-Gauss-Bonnet set-up. \\ 
The statistical anisotropy during slow-roll inflation can be attributed to the violation of the 
spatial de-Sitter symmetry resulting into
directional dependence of the power spectrum given by,
\begin{equation}
 P(\overrightarrow{{\bf k}})=P(k) \left[1+ g^{*}(\hat{{\bf k}}.\overrightarrow{{\bf n}})^2\right] 
\end{equation}
where $\hat{k}$ is the unit vector along the direction of the wavenumber vector $\overrightarrow{{\bf k}}$,
$\overrightarrow{{\bf n}}$ is the vector that breaks the rotational invariance which in the
present case is taken in the direction of x-axis and $g^{*}$ denotes anisotropy in the power spectrum.
The current Planck data admits both positive and negative values of $g^{*}$ 
and places the upper bound to be $g^{*}<0.23 \times 10^{-2}$ \cite{Planck-2015}.
The present work leaves the scope of both negative and positive $g^{*}$ because the sign of
measure of anisotropy depends on given choice of the ratio $\displaystyle\frac{\dot{v_2}}{\dot{v_1}}$.\\
With quadratic forms of inflaton potential $V(\phi)$ and Gauss-Bonnet coupling function $\xi(\phi)$,
we can compare the
observational data of scalar spectral index $n_s$ and 
tensor-to-scalar ratio $r$ (using Planck $+$ WMAP $+$ Baryon Acoustic Oscillations (BAO)+highL) 
with the respective theoretical values of $n_s$ and $r$  which will include inputs namely
$g^{*}$, positive Gauss-Bonnet coupling parameter $\xi_0$ and the e-folding number $N$. 
In order to investigate the role of the non-abelian vector field in presence of
higher curvature corrections, one can consider 
different values of $g^{*}$, $\xi_0$ and large value of $N$ ($>60$)
while satisfying the condition $m^2 \xi_0<<1$ 
such that comparison with the observational data will put 
constraint on both $\xi_0$ and $g^{*}$ ( and hence on $c$)
for the chosen forms of $V(\phi)$ and $\xi(\phi)$.
This analysis can help us determine allowed ranges for $g^{*}$, $\xi_0$ and also
compare with the theoretically predicted range $\xi_0 < 10^{7}$ 
so as to ascertain the contribution of the Gauss-Bonnet term.
If the obtained values of $n_s$ and $r$
lie in the `` sweet spot " of the observations, role of non-abelian vector models
as the source of anisotropy as well as presence of Gauss-Bonnet corrections can be established
during slow-roll inflation.

\section*{Acknowledgements}
I am thankful to Jiro Soda and Narayan Banerjee 
for valuable suggestions and fruitful discussions at various stages of the work. 
I would like to thank Claus Laemmerzahl in ZARM where part of the work had been completed.

\bibliographystyle{JHEP}

\end{document}